%% file: kopeliovich_IS.tex
\documentclass{aip-cp}

\usepackage[numbers]{natbib}
\usepackage{rotating}
\usepackage{graphicx}

\def\im{{\rm Im}}
\def\re{{\rm Re}}

 \newcommand\la{\langle}
 \newcommand\ra{\rangle}
 \newcommand\beq{\begin{equation}}
 
 \newcommand\eeq{\end{equation}}
 \newcommand\beqn{\begin{eqnarray}}
 \newcommand\eeqn{\end{eqnarray}}

\def\GeV{\,\mbox{GeV}}

\begin{document}

\title{Leading Neutrons From Polarized\\ 
Proton-Nucleus Collisions}

\author{B. Z. Kopeliovich\corref{cor1}}
\author{I. K. Potashnikova}
\eaddress{irina.potashnikova@usm.cl}
\author{Ivan Schmidt}
\eaddress{ivan.schmidt@usm.cl}

\affil{Departamento de F\'{\i}sica,
Universidad T\'ecnica Federico Santa Mar\'{\i}a;\\ and
Centro Cient\'ifico-Tecnol\'ogico de Valpara\'iso;
Casilla 110-V, Valpara\'iso, Chile}
\corresp[cor1]{Corresponding author: boris.kopeliovich@usm.cl}

\maketitle

\begin{abstract}
Leading neutron production on protons is known to be subject to strong absorptive corrections, which have been under debate for a long time. On nuclear targets these corrections are significantly enhanced and push the partial cross sections of neutron production to the very periphery of the nucleus. As a result, the $A$-dependences of inclusive and diffractive neutron production turn out to be similar. 
The mechanism of $\pi$-$a_1$ interference, which successfully explained the observed single-spin asymmetry of neutrons in polarized pp interactions, is extended here to polarized $pA$ collisions. Corrected for nuclear effects it explains quite well the magnitude and sign of the asymmetry $A_N$ observed in inelastic events, resulting in a violent break up of the nucleus. However the excessive magnitude of $A_N$ observed in the diffractive sample, remains a challenge.
\end{abstract}

\section{NEUTRON PRODUCTION IN THE VICINITY OF PION POLE}

Pions are known to have a large coupling with nucleons, therefore
pion exchange is important in the processes with iso-spin flip, like  $p\to n$. Measurements with polarized proton beams supply more detailed information about the interaction dynamic.

\begin{figure}[htb]
\centerline{
  \scalebox{0.35}{\includegraphics{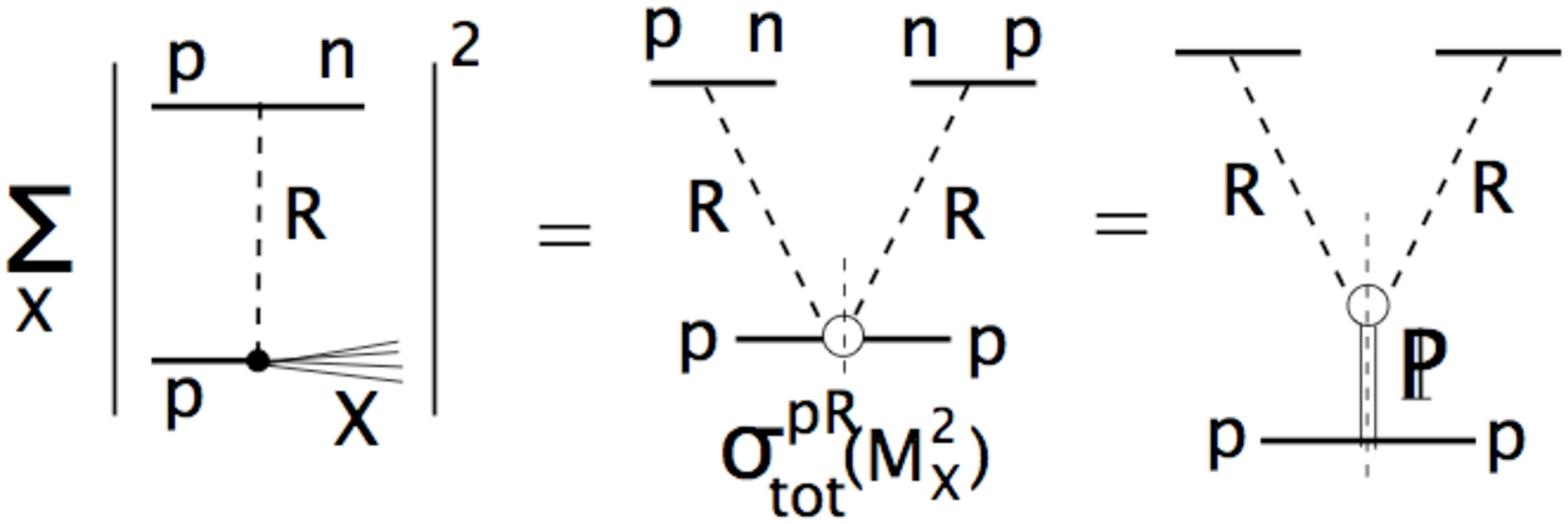}}}
\caption{\label{fig:3R} The cross section of neutron production is proportional to the total Reggeon-proton cross section, which at large $M_X^2$ is dominated by Pomeron exchange.}
 \end{figure}

The process $p+p(A)\to n+X$,
with a large fractional light-cone momentum, $z=p^+_n/p^+_p$,
of neutrons produced in the proton beam direction, is known to
be related to the iso-vector Reggeons ($\pi,\ \rho,\ a_2,\ a_1$, etc.)
\cite{kpss},
as is illustrated in figure~\ref{fig:3R}, where
the amplitude squared and summed over the final states $X$ (at a fixed invariant mass $M_X$) is expressed via the Reggeon-proton total cross section at the c.m. energy $M_X$. 

At  high energies of colliders (RHIC, LHC) $M_X^2=s(1-z)$ is so large (except  at inaccessibly small $1-z$), that the cross section $\sigma^{pR}_{tot}(M_X^2)$ is dominated by Pomeron exchange, as is illustrated in figure~\ref{fig:3R}. 

The couplings of iso-vector Reggeons with natural parity ($\rho$, $a_2$) to the proton  are known to be predominantly spin-flip \cite{kane}, so they can be neglected, because we are interested here in small transverse momenta of neutrons, $p_T\to0$. 
Only unnatural parity Reggeons ($\pi$, $a_1$), having large spin non-flip couplings contribute in the forward direction.

The pion term reads \cite{kpss},
\beq
\left.z\,\frac{d\sigma^B(pA\to nX)}{dz\,dq_T^2}\right|_{\pi}=
f_{\pi/p}(z,q_T,q_L)\,
\sigma^{\pi^+ A}_{tot}(s'=M_X^2),
\label{100}
 \eeq
where $f_{\pi/p}(z,q_T,q_L)$ is the pion flux in the proton,
 \beq
f_{\pi/p}(z,q_T,q_L)=
|t|\,G_{\pi^+pn}^2(t)\left|\eta_\pi(t)\right|^2
\left(\frac{\alpha_\pi^\prime}{8}\right)^2 
(1-z)^{1-2\alpha_\pi(t)}.
\label{120}
 \eeq
Here $q_T$ is the neutron transverse momentum; 
$
q_L=m_N(1-z)/\sqrt{z}
$; 
$
-t=q_L^2+q_T^2/z
$;  and $G_{\pi^+pn}(t)$ is the effective $\pi$-$N$ vertex function \cite{kpp,kpss}.

The amplitude of the process includes both non-flip and spin-flip terms \cite{kpss},
\beq
A^B_{p\to n}(\vec q,z)=
\bar\xi_n\left[\sigma_3\, q_L+
\frac{1}{\sqrt{z}}\,
\vec\sigma\cdot\vec q_T\right]\xi_p\,
\phi^B(q_T,z),
\label{900}
\eeq
The superscript $B$ here and in (\ref{100}) means that this is the Born approximation, and then the higher order re-scattering corrections, usually called absorption, are ignored. The procedure of inclusion of the absorptive corrections on the amplitude level was developed in \cite{kpss,kpps}. First, the amplitude is Fourier transformed to impact parameter representation, where the absorptive effect is just a multiplicative suppression factor. Then the absorption corrected amplitude is Fourier transformed back to momentum representation. The effects of absorption turn out to be different for the non-flip and spin-flip terms in the amplitude.
Here we skip the details of this procedure, which was described in  detail in \cite{kpss,kpps,pi-pi}. In accordance with the spin structure of the amplitude, the resulting cross section gets contributions from the non-flip and spin-flip terms,
\beq
\left.z\,\frac{d\sigma(pp\to nX)}{dz\,dq_T^2}\right|_{\pi}=
\sigma_0^\pi(z,q_T) + \sigma_s^\pi(z,q_T)\,,
\label{160}
 \eeq
 which were calculated in \cite{kpss}.

\section{Leading neutrons from \MakeLowercase{p}A collisions}

A natural extension of equation (\ref{100}) to nuclear targets has the form,
\beq
\left.z\,\frac{d\sigma(pA\to nX)}{dz\,dq_T^2\,d^2b_A}\right|_{\pi}=
f_{\pi/p}(z,q_T)\,\frac{d\sigma^{\pi A}_{tot}(M_X^2)}{d^2b_A}\,
 S_{NA}(b_A),
\label{580}
\eeq
where $b_A$ is the impact parameter of $pA$ collision; $S_{NA}(b_A)$ is the additional nuclear absorption factor described below.

This expression can be interpreted as interaction of the projectile Fock component $|\pi^+n\ra$  with the target, sharing the proton light-cone momentum in fractions $z$ and $(1-z)$ respectively. While the pion interacts inelastically with the target, the spectator neutron has to remain intact, i.e. has to survive propagation through the nucleus.

The partial total pion-nucleus cross section in Eq.~(\ref{580}) can be evaluated in the Glauber approximation,
\beqn
\left.\frac{\sigma^{\pi A}_{tot}(M_X^2)}{d^2b_A}\right|_{Gl} 
\approx
2\re\left\{1-
e^{-{1\over2}\,\sigma^{\pi N}_{tot}(M_X^2)\,
T_A(b_A)}\right\}\, ,
\label{620}
\eeqn
 where 
$
T_A(b_A) = \int_{-\infty}^\infty d\zeta\,\rho_A(b_A,\zeta)\ ,
$
 is the nuclear thickness function; $\rho_A(b_A,\zeta)$
 is the nuclear density.

Correspondingly, the neutron survival factor in (\ref{580}) reads,
\beq
S_{NA}(b_A)\bigr|_{Gl}\approx
\frac{e^{-\sigma^{nN}_{in}(zs)T_A(b_A)}-e^{-\sigma^{pN}_{in}(s)T_A(b_A)}}
{T_A(b_A)[\sigma^{pN}_{in}(s)-\sigma^{nN}_{in}(zs)]}
\approx 
e^{-\sigma^{NN}_{in}(s)T_A(b_A)}.
\label{800}
\eeq

The Gribov corrections \cite{gribov69} are also included implicitly in all our calculations. They are known to make the nuclear matter more transparent for propagation of  hadrons \cite{zkl,mine} and affect both factors in (\ref{580}), suppressing $\sigma^{\pi N}_{tot}$ and increasing $S_{NA}(b_A)$. We calculate the Gribov corrections to all orders of multiple interactions by employing the dipole representation, as is described in  \cite{zkl,mine,kps,gribov85}. 

The specifics of  measurements of forward neutrons in the PHENIX experiment was supplemented by the Beam-Beam Counters (BBC), detecting charged particles in two pseudo-rapidity intervals $3.0<|\eta|<3.9$. The results of measurements of forward neutrons are
presented for different samples of events: (i) inclusive neutron production;
(ii) neutrons associated with multi-particle production (one or both BBCs are fired);
(iii) "might be" rapidity gap diffractive production of neutrons (both BBCs are vetoed).

If BBC are fired by multi-particle production, one should modify equation (\ref{580}) replacing,
\beq
2\left[1-e^{-{1\over2}\sigma^{\pi N}_{tot}T_A(b_A)}\right]\ 
\Rightarrow \ 
1-e^{-\sigma^{\pi N}_{in}T_A(b_A)}
\label{885}
\eeq

If BBC are vetoed, one can think about the diffractive channels $p+A\to n\pi^++A^*$, which escape detection in the BBC. Then instead of the total $\pi A$ cross section in (\ref{580}), one should be restricted to the  elastic and quasi-elastic channels,  replacing, 
\beq
2\left[1-e^{-{1\over2}\sigma^{\pi N}_{tot}T_A(b_A)}\right]\ 
\Rightarrow \ 
\left[1-e^{-{1\over2}\sigma^{\pi N}_{tot}T_A(b_A)}\right]^2
+\sigma^{\pi N}_{el}\,T_A(b_A)\,e^{-\sigma^{\pi N}_{in}T_A(b_A)}
\label{890}
\eeq

\begin{figure}[htb]
\centerline{
  \scalebox{0.35}{\includegraphics{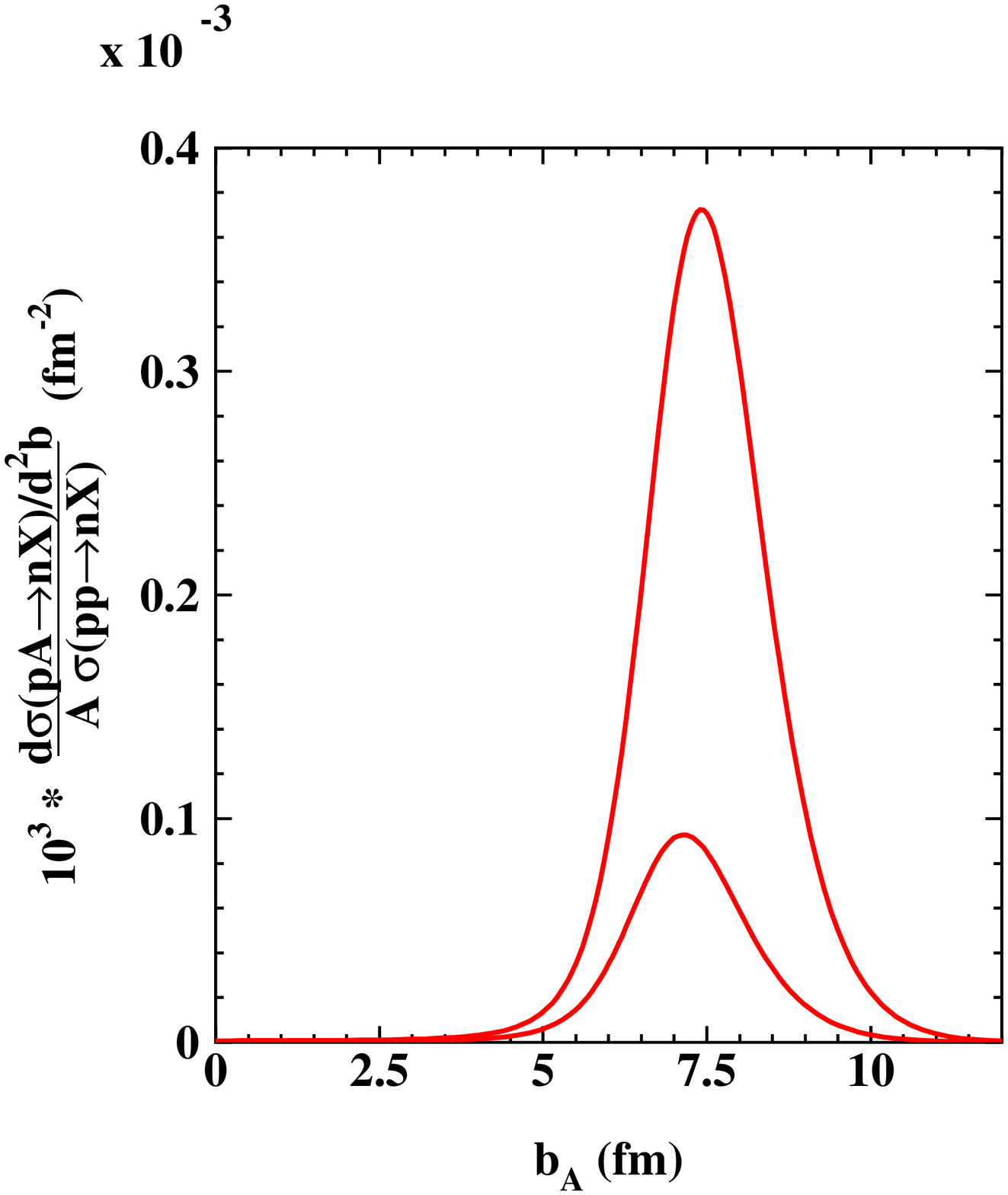}}
  \hspace{1cm}
    \scalebox{0.35}{\includegraphics{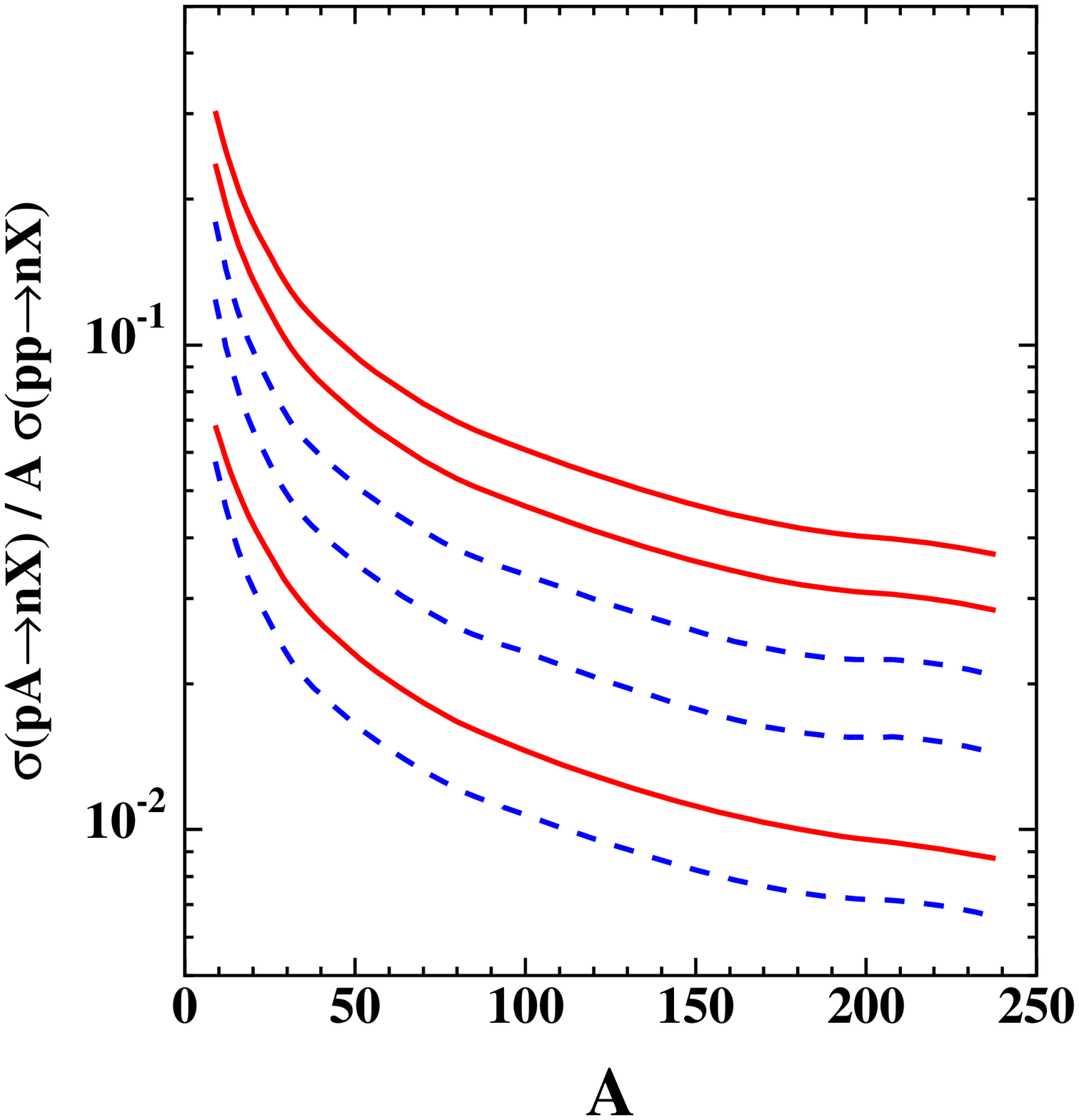}}
    }
\caption{\label{fig:sig} {\sl Left:} Partial cross sections for inclusive (upper curve) and diffractive (bottom curve) neutron production in $pAu$ collisions at $\sqrt{s}=200\GeV$.
{\sl Right:}  $b_A$-integrated cross sections, normalized by the $pp$ cross section of leading neutron production. The solid and dashed curves correspond to both $pAu$ and $pPb$ collisions at $\sqrt{s}=200\GeV$ and $5000\GeV$ respectively. The three curves in each set from top to bottom correspond to inclusive, inelastic and diffractive neutron production, respectively.}
 \end{figure}

The results of Glauber model calculations, including Gribov corrections, of the partial cross sections for $pAu\to nX$ at $\sqrt{s}=200\GeV$, are plotted in the left pane of figure~\ref{fig:sig}.
The upper and lower curves correspond to inclusive and diffractive production respectively.
Apparently, strong absorption allows neutron production only on nuclear periphery.

\section{Single-spin asymmetry}
\subsection{\boldmath$A_N$ in polarized $pp\to nX$}

Both spin amplitudes in equation~(\ref{900}) have the same phase factor, $\eta_\pi(t)=i-ctg\left[\pi\alpha_\pi(t)/2\right]$.
Therefore, in spite of the presence of both spin-flip and non-flip amplitudes, no single-spin asymmetry associated with pion exchange is possible in the Born approximation. Even the inclusion of absorptive corrections leave the spin effects miserably small \cite{kpss-spin} compared to data \cite{phenix-pp1,phenix-pp2}. 

A plausible candidate to produce a sizeable spin asymmetry is $a_1$ meson exchange,
since $a_1$ can be produced by pions diffractively. However, this axial-vector resonance 
is hardly visible in diffractive channels $\pi+p\to 3\pi+p$, which are dominated by $\pi\rho$
in the $1^+S$ wave. The $\pi\rho$ invariant mass distribution forms a pronounced narrow peak at $M_{\pi\rho}\approx m_{a_1}$ (due to the Deck effect \cite{deck}). Although in the dispersion relation for the amplitude this channel corresponds to a cut, it can be represented as an effective pole $\tilde a_1$ \cite{kpss-spin}.
In the crossed channel, $\pi\rho$ exchange corresponds to a Regge cut, with known intercept and slope of the Regge trajectory \cite{kpss-spin}.

The expression for the single-spin asymmetry arising from $\pi$-$\tilde a_1$ interference was derived in \cite{kpss-spin} and has the form,
\beq
A_N^{(\pi-\tilde a_1)}(q_T,z) =
q_T\,\frac{4m_N\,q_L}{|t|^{3/2}}\,
(1-z)^{\alpha_\pi(t)-\alpha_{\tilde a_1}(t)}\,
\frac{\im\,\eta_\pi^*(t)\,\eta_{\tilde a_1}(t)}
{\left|\eta_\pi(t)\right|^2}\,
\left(\frac{d\sigma_{\pi p\to \tilde a_1p}(M_X^2)/dt|_{t=0}}
{d\sigma_{\pi p\to\pi p}(M_X^2)/dt|_{t=0}}\right)^{1/2}
\frac{g_{\tilde a_1^+pn}}{g_{\pi^+pn}}.
\label{920}
\eeq
The trajectory of the $\pi\rho$ Regge cut and the phase factor $\eta_{\tilde a_1}(t)$ are known from Regge phenomenology.
The ${\tilde a_1}NN$ coupling was evaluated in \cite{kpss-spin}, based on PCAC and the 
second Weinberg sum rule, where the spectral functions of the vector and axial
currents are represented by the $\rho$ and the effective ${\tilde a_1}$ poles respectively.
This leads to the following relations between the couplings,
\beq
\frac{g_{\tilde a_1 NN}}{g_{\pi NN}}=
\frac{m_{\tilde a_1}^2\,f_\pi}{2m_N\,f_\rho}\approx {1\over2},
\label{340}
\eeq
where $f_\pi=0.93m_\pi$ is the pion decay coupling;
$f_\rho=\sqrt{2}m_\rho^2/\gamma_\rho$, 
and $\gamma_\rho$ is the universal coupling, $\gamma_\rho^2/4\pi=2.4$.

\begin{figure}[htb]
\centerline{
 {\includegraphics[height=6.5cm]{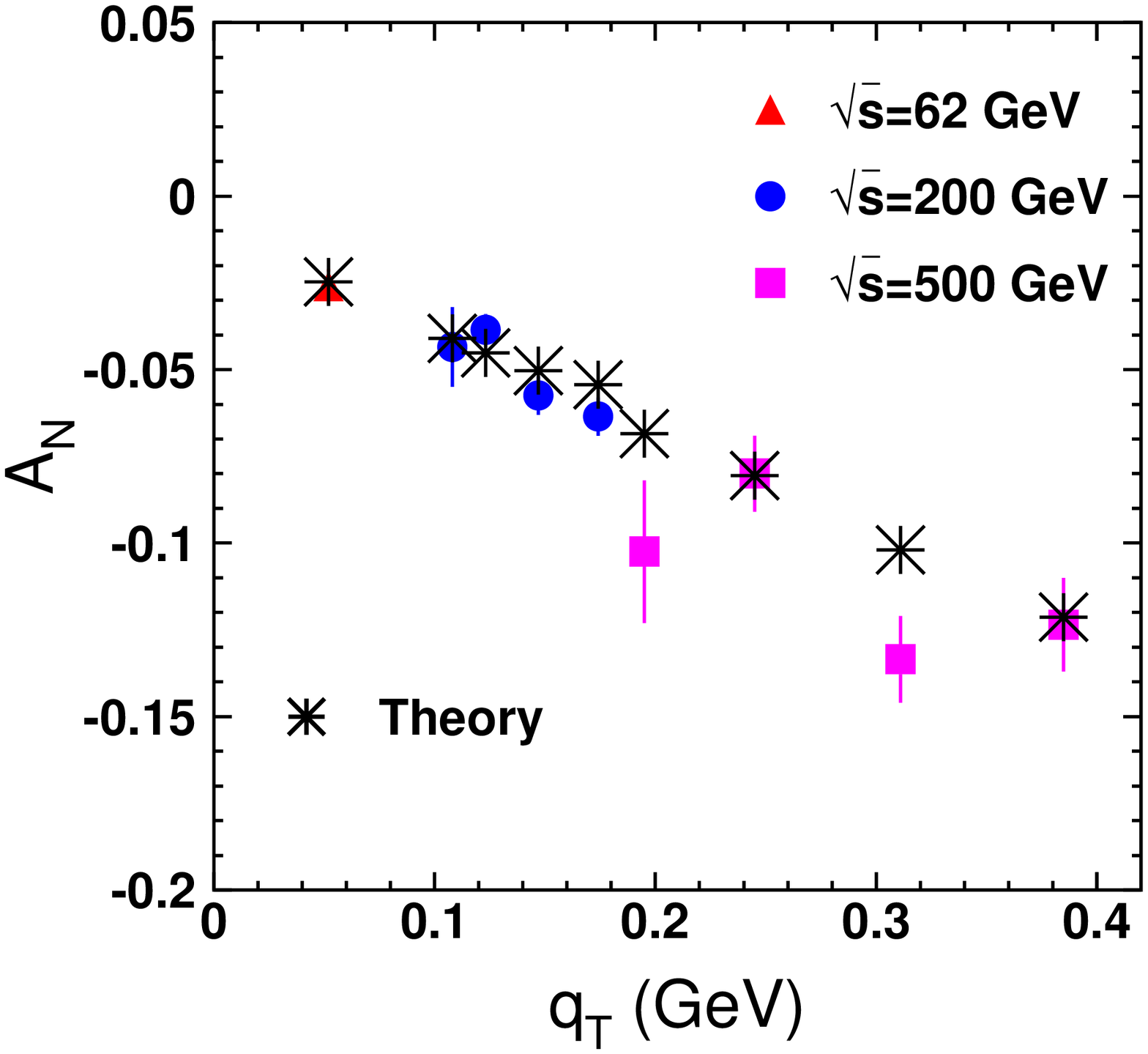}}
 \hspace*{-2cm}
 {\includegraphics[height=7.5cm]{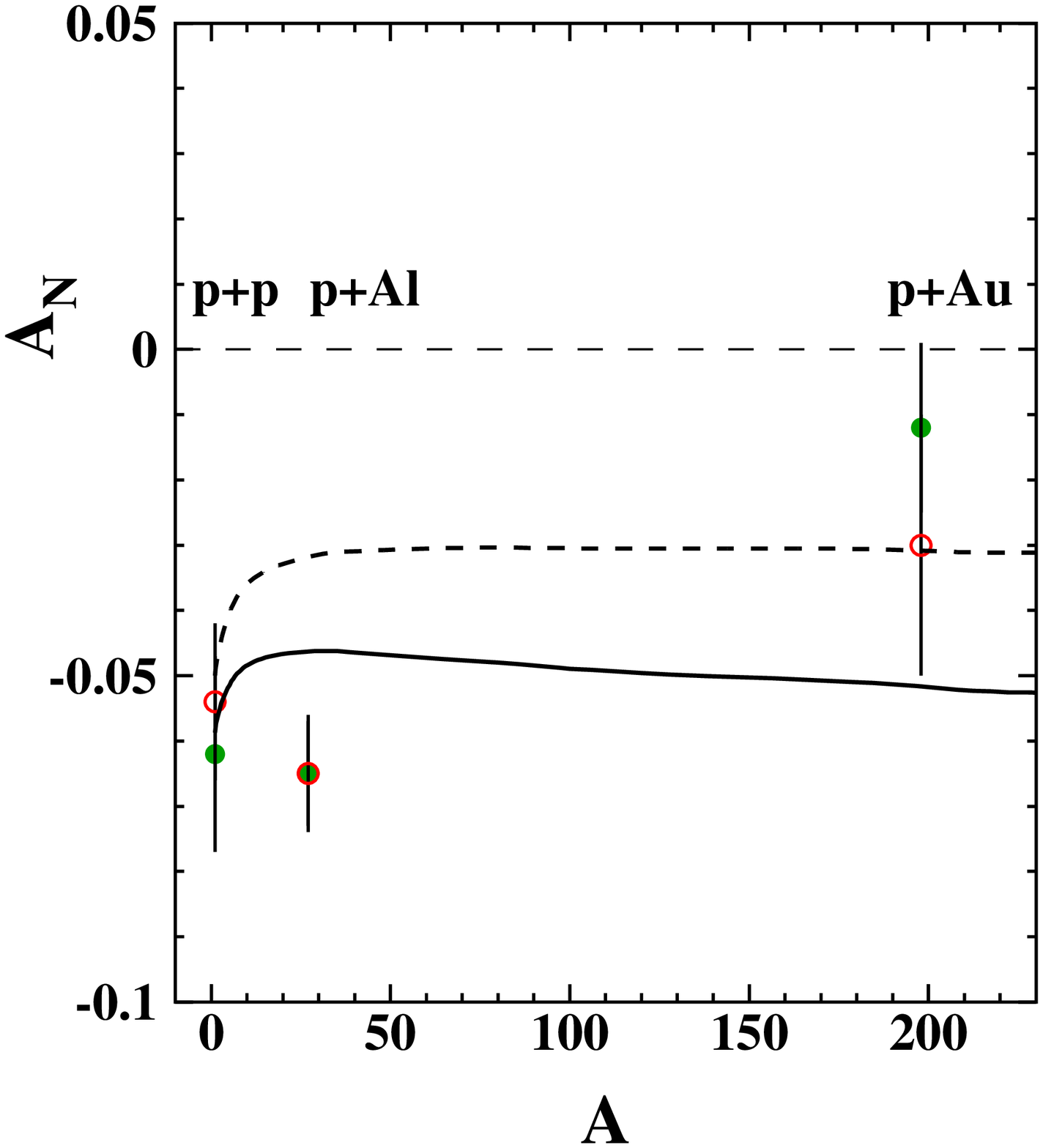}}
 }
\caption{\label{fig:AN} {\sl Left:} Single-spin asymmetry $A_N$ in polarized $pp\to nX$ at $\sqrt{s}=200\GeV$ vs neutron transverse momentum $q_T$. Data points from \cite{phenix-pp1,phenix-pp2} are self-explained, the stars show the results of the parameter-free calculations with proper kinematics \cite{kpss-spin}.
{\sl Right:} $A_N$ in polarized $pA\to nX$ vs $A$ at $\la q_T\ra=0.115\GeV$ and $\la z\ra = 0.75$. Full and open data points correspond to events with either both BBCs fired, or only one of them fired in the nuclear direction, respectively \cite{bazil1,itaru,bazil2}. An attempt to model these two classes of events is presented by solid and dashed curves  (see text).
}
 \end{figure}

The parameter-free calculations agree with the PHENIX data \cite{phenix-pp1,phenix-pp2}, as is depicted in the left pane of figure~\ref{fig:AN}.

\subsection{\boldmath$A_N$ in polarized $pA\to nX$}

The single-spin asymmetry on a nuclear target due to $\pi$-$\tilde a_1$ interference can be calculated with a modified equation (\ref{920}), replacing 
\beq
\frac{d\sigma_{\pi p\to \tilde a_1p}(M_X^2)/dt|_{t=0}}
{d\sigma_{\pi p\to\pi p}(M_X^2)/dt|_{t=0}}
\,\Rightarrow\,
\frac{d\sigma_{\pi A\to \tilde a_1A}(M_X^2)/dt|_{t=0}}
{d\sigma_{\pi A\to\pi A}(M_X^2)/dt|_{t=0}}
\label{500}
\eeq

This replacement leads to the single-spin asymmetry, which can be presented in the form,
\beq
A_N^{pA\to nX} =
A_N^{pp\to nX}\times \frac{R_1}{R_2}\,R_3,
\label{520}
\eeq 
where
\beq
R_1= \frac{1}{\sigma^{\rho p}_{tot}}\int d^2b\, e^{-{1\over2}\sigma^{\pi p}_{tot}T_A(b)}
\left[1-e^{-{1\over2}\sigma^{\rho p}_{tot}T_A(b)}\right]
 e^{-{1\over2}\sigma^{pp}_{tot}T_A(b)},
 \label{540}
 \eeq
 represents the nuclear effects for coherent $\pi A\to \pi\rho A$;
\beq
 R_{2}= \frac{2}{\sigma^{\pi p}_{tot}}\int d^2b\,
\left[1-e^{-{1\over2}\sigma^{\pi p}_{tot}T_A(b)}\right]
 e^{-{1\over2}\sigma^{pp}_{tot}T_A(b)},
 \label{560}
 \eeq
 is the nuclear factor for the denominator $\pi A\to \pi A$.
 
The factor $R_3=\sigma^{\pi A}_{tot}/\sigma^{\pi A}_{in}$ takes into account the specific 
trigger on inelastic collisions, which fire both BBCs, otherwise it is fixed at $R_3=1$. The results corresponding to these two choices are plotted in the right pane of figure~\ref{fig:AN}, by solid and dotted curves respectively.
The difference of these two results reflects the big uncertainly in the physical interpretation of events with fired of vetoed BBCs. This can be improved by applying a detailed Monte-Carlo modelling. Nevertheless, our calculations reproduce  reasonably well the results of measurements \cite{bazil1,itaru,bazil2}.

\section{Summary}

The previously developed methods of calculation of the cross section of leading neutron production in $pp$ collisions, are extended to nuclear targets. The nuclear absorptive corrections are so strong that push the partial cross sections of leading neutron production to the very periphery of the nucleus. As a result, the $A$-dependences of inclusive and diffractive neutron production turn out to be similar. 

The mechanism of $\pi$-$a_1$ interference, which successfully explained the observed single-spin asymmetry in polarized reaction $pp\to nX$, is extended to polarized $pA$ collisions. Corrected for nuclear effects, it explains quite well the observed asymmetry $A_N$ in inelastic events, when the nucleus violently breaks up. However, the large value and opposite sign of $A_N$ observed in the diffractive sample, remains a challenge.

\section{ACKNOWLEDGMENTS}
We are thankful to Alexander Bazilevsky, Itaru Nakagawa and Minjung Kim 
for providing us with preliminary data and details of the measurements, as well as
for numerous informative discussions.
This work was supported in part
by Fondecyt (Chile) grants 1130543, 1130549 and 1140842,
by Proyecto Basal FB 0821 (Chile),
and by CONICYT grant  PIA ACT1406 (Chile).

\input{kopeliovich-ref}

\end{document}

%% file: kopeliovich-ref.tex
\nocite{*}
\bibliographystyle{aipnum-cp}%
\bibliography{sample}%